\documentstyle[12pt]{article}
\input{psfig}
\newcommand{\beq}{\begin{equation}}
\newcommand{\eeq}{\end{equation}}
\def\bea{\begin{eqnarray}}
\def\eea{\end{eqnarray}}
\newcommand{\ba}{\begin{array}}
\newcommand{\ea}{\end{array}}

\def\et{{\it et al.}}
\def\cp{\Delta_{CP}} 
\def\aw{\alpha_W}
\def\w{\omega}

\begin{document}

\begin{titlepage}
\begin{flushright}
CERN-TH/96-301 \\
hep-ph/9611227
\end{flushright}
\vspace{2cm}
\centerline{\Large\bf Neutral Heavy Leptons and Electroweak Baryogenesis}
%\vspace{10pt}
%\centerline{\Large\bf }
\vspace{24pt}
\centerline{\large P. Hern\'andez and N. Rius
\footnote{On leave of absence from Departament de F\'{\i}sica Te\`orica, 
Universitat de Val\`encia and IFIC, Val\`encia, Spain.}}
\centerline{\it Theory Division, CERN}
\centerline{\it CH-1211 Geneva 23, Switzerland}

\vspace{4cm}
\begin{abstract}
We investigate the possibility that 
baryogenesis occurs during the weak phase transition in a
minimal extension of the Standard Model which contains
extra neutral leptons and conserves total lepton number. 
The necessary CP-violating phases appear in the leptonic 
Yukawa couplings.
We compute the CP-asymmetries in both the neutral and
the charged lepton fluxes reflected in the unbroken phase.
Using present experimental bounds on the mixing angles 
and Standard Model estimates for the parameters related
to the scalar potential, we conclude that it seems unlikely to 
produce the observed baryon to entropy ratio within this 
kind of models. However, we comment on the possibility that the
constraints on the mixings might be naturally relaxed due to small finite 
temperature effects. 
\end{abstract}
\vspace{5cm}
CERN-TH/96-301\\
October 1996
\end{titlepage}
%\

\renewcommand{\thefootnote}{\arabic{footnote}}
\setcounter{footnote}{0}

\section{Introduction}

The baryon number to entropy ratio in the observed part of the Universe 
is required to be 
$n_B / s \sim (4$--$6)\times 10^{-11}$ 
by nucleosynthesis constraints \cite{nuc}. 
In 1967, Sakharov \cite{sak} established the three basic requirements 
for obtaining this baryon asymmetry as a result of particle 
interactions in the early universe:  
a) Baryon number violation,
b) C and CP violation, 
c) departure from thermal equilibrium.
These conditions may be fulfilled at weak scale temperatures 
\cite{krs}, if the electroweak phase transition is first order.   

In a strongly first order electroweak transition, bubbles of the true ground state (broken phase) nucleate 
and expand until they fill the Universe; local departure from 
thermal equilibrium occurs in the vicinity of the expanding 
bubble walls. The other two Sakharov conditions are also 
satisfied, since C and CP are known to be violated by the electroweak 
interactions and anomalous baryon number violation is fast at high 
temperatures in the symmetric phase.
As a bubble expands, particles in the unbroken phase will reflect off
the advancing wall. CP-violating interactions result in 
a different reflection probability for fermions with a given chirality  and  the corresponding
 antifermions,  
leading to a CP asymmetry in the reflected chiral number flux \cite{ckn}.  
In the symmetric phase, anomalous $B+L$ violating interactions are
in thermal equilibrium and the reflected current induces a net 
baryon number.
An important survival requirement for the produced baryon asymmetry 
is that the sphaleron processes inside the bubble are slow enough 
 and this in turn is directly related to the strength of the phase transition.

In principle the Standard Model (SM) contains all the necessary 
ingredients for electroweak baryogenesis, but it has two problems:
the CP asymmetry induced by  the Kobayashi-Maskawa phase is 
far too small to account for the observed $n_B / s$ ratio \cite{b2,hs}, 
and the phase transition appears too weakly first order for the 
Higgs mass experimentally allowed \cite{pt}. 
However, these two problems may be
absent in several 
simple extensions of the SM, which contain additional 
sources of CP violation and more scalars than the SM.
The larger parameter space in the scalar sector allows
for a stronger first-order phase transition without such a light Higgs
\cite{ah}. 
Several such possibilities have been considered in the literature: two Higgs models
with a strong CP phase \cite{cknh}--\cite{cline}, 
heavy Majorana neutrinos \cite{cknn}, and supersymmetric models \cite{hn}. 

In the present paper, we consider models with an extended lepton sector, 
which conserves total lepton number. 
The model provides a viable alternative to the 
see-saw mechanism for explaining the lightness of the known 
neutrinos, in those extensions of the SM where there are no scalars 
carrying lepton number, which could generate Majorana masses. They arise 
in several contexts, such as GUTs \cite{ww} and
E(6) superstring-inspired models \cite{valle}. Similar patterns of lepton
masses have also been obtained in the context of models of 
Extended Technicolor with a GIM mechanism \cite{lisa}.  

The relevant features for baryogenesis are twofold. First, 
the lepton Yukawa interactions contain additional CP-violating phases, 
which can lead to a much larger CP asymmetry than the 
CKM phase in the SM. 
In contrast to the models with Majorana neutrinos considered in \cite{ckn}, 
the CP-violating effects in this case are not suppressed by 
the light neutrino masses. Second, the presence of an additional
singlet scalar may help in getting a stronger first-order phase transition.

An interesting issue that may be relevant in this type of models is whether 
finite temperature corrections can produce an enhancement of the CP asymmetry,
as was  found in the first detailed calculation of this quantity in the 
SM \cite{fs}
\footnote{To be more precise, the typical suppression with quark masses 
expected
in a flavour-blind CP-violating process was not found, while the suppression
with the CP-conserving angles was explicit.}.
In \cite{b2}\cite{hs} it was shown that this enhancement disappears in the SM 
when one properly includes the incoherence effects induced by the 
interaction of the quarks with the plasma. However, the question remained that 
if the particles involved were much more weakly interacting, as leptons 
instead of quarks, maybe this enhancement would be at work. 

In section 2 we describe the model, and the order of 
magnitude estimates of the CP-violating asymmetries are obtained
in section 3. We find that the leading effects are different for the 
reflection of the neutral and the charged leptons. Naively the 
later is smaller since it is suppressed by the charged lepton masses;
however, an enhancement of the type of ref. \cite{fs} could 
imply no such suppression.
In section 4
we compute the contribution to the asymmetry due to the reflection of the 
neutral leptons, which turns out to be the leading effect, as expected. 
In close analogy with the SM case, we consider in section 5 the lepton 
asymmetry generated by the charged lepton reflection on the bubble wall. 
As we will see, no enhancement with respect to the naive estimate is found. 
In section 6 we compute the baryon number induced by the CP asymmetries 
in the neutral sector, and we conclude in section 7.

\section{The Model}

The phenomenology of this type of models has been extensively studied in
\cite{berna}\cite{lisa}. Here we briefly describe the essential features
relevant for baryogenesis. 

The gauge group is the standard $SU(2) \times U(1)$, with minimal quark sector.
The lepton sector is extended with two
electroweak singlet two-component leptons in each generation,  
i.e., 
\beq
\Psi_L^i  =  \left( \ba{c}
\nu_L^i  \\ 
e_L^i
\ea \right) , \; e_R^i, \;  \nu_R^i, \; s_L^i.
\eeq
Unlike the minimal standard model, total lepton number 
conservation is not an automatic symmetry. It has to 
be imposed, and it restricts the form of the Yukawa terms that lead 
to the neutral fermion masses, while the Yukawa terms
involving charged leptons are completely standard:  
\beq
{\cal L}_Y = \bar{\Psi}_L f H e_R   +  
  \bar{\Psi}_L f_D H \nu_R  + \bar{s}_L f_S \sigma \nu_R + h.c.
%  -V(|\sigma|^2, |H|^2),
\label{lag} 
\eeq
where $f_i$ are the Yukawa matrices, $H$ is the standard Higgs doublet
and $\sigma$ is a new singlet scalar field. Due to the presence of $\sigma$, 
the weak phase transition can be quite strongly first order for 
a significant range of parameters \cite{ah}.

For simplicity, we will assume that the singlet $\sigma$ acquires a vacuum 
expectation value at the same scale as the standard Higgs doublet, 
and the theory undergoes a single phase transition at a critical 
temperature near the weak scale. Although this does not need to be
the case, we expect that the CP asymmetry will not 
depend much on this choice (at least if power counting arguments give 
a correct estimate).
Then, in the broken phase the lepton mass terms are
\beq
\bar{e}_R m e_L  +
\bar{\nu}_R D \nu_L +
\bar{\nu}_R S s_L  + h.c., 
\label{mass} 
\eeq
where 
\beq
m = f^\dagger  v  \ \ \,    
D = f_D^\dagger  v \ \ \,
S = f_S^\dagger u 
%\label{} 
\eeq
$v$ and $u$ being the vevs of the doublet and singlet scalar fields,
respectively.

The structure of the neutral sector mass matrix (\ref{mass}) ensures the 
existence of three massless Weyl neutrinos $\nu^0$, 
regardless of the relative value of $D$ and $S$.
The other six Weyl fermions pair up into three heavy neutral Dirac 
fermions $n$, whose masses are essentially determined by $S$.
Note that in this model the ratio $D/S$ is expected to be less constrained 
than the corresponding parameter of any model that invokes the see-saw
mechanism to understand the smallness of neutrino masses. 
This is because in such models the neutrino masses are only
suppressed by $D/S$, which is therefore very constrained. 
In the present case, this ratio is not related to the light neutrino masses.
Nevertheless, we will see in the next section that it can also be constrained 
from the strong bounds on charged lepton mixing. 

The mass matrix can be diagonalized by multiplying on the right 
and the left by unitary matrices:
\beq
\left (
\ba{cc}
V & 0  \\ 
0 & 0
\ea \right)
\left (
\ba{cc}
D & S  \\ 
0 & 0
\ea \right)
U = 
\left (
\ba{cc}
0 & 0  \\ 
0 & M
\ea \right)
\eeq
with
\beq
\left (
\ba{c}
\nu_L  \\ s_L
\ea \right)
 = U \left (
\ba{c}
\nu^0 \\ n_L
\ea \right )
\label{cb}
\eeq
where $\nu^0$ are the massless neutrinos and $n$ are 
the neutral heavy leptons (NHL).

The unitary matrix $V$ diagonalizes $D D^\dagger + S S^\dagger$
to give $M^2$. The unitary matrix $U$ can be written as
\beq
U=
\left(
\ba{cc}
K_L & K_H \\
K_{SL} & K_{SH} \\
\ea
\right)
\label{u}
\eeq
where $K_L, K_{SH} \sim 1 - {\cal O}[(D/S)^2]$ and 
$K_H, K_{SL} \sim {\cal O}(D/S)$.

Using (\ref{cb}) we get the following form 
for the charged current leptonic weak interaction:
\beq
{\cal L}_{cc} = {g \over 2} W^\mu \sum_{a=1}^3 \sum_{\alpha=1}^6
\bar{e}_a \gamma_\mu L 
\left[ K_L  \nu^0 + K_H n_L \right]_{a \alpha}  
+ h.c.
\label{cc}
\eeq 
One can see that the charged current coupling of the mass eigenstates 
charged leptons to the massless as well as the heavy neutrinos is
non-trivial, making possible the violation of individual lepton 
numbers $L_e, L_\mu$ and $L_\tau$. 

Due to the admixture of fermions of different weak isospin,
there is no GIM mechanism in the neutral fermion couplings to the 
$Z$ boson, which are given by
\beq
{\cal L}_{nc} = {g \over 2 c_w} Z^\mu \sum_{\alpha, \beta}   
\bar{N}_\alpha \gamma_\mu L P_{\alpha \beta} N_\beta,    
\label{nc}
\eeq
where $N_\alpha = (\nu^0_a, n_a)$, and 
$P=K_L^\dagger K_L + K_H^\dagger K_H$ 
is in general a non-diagonal projection matrix.
The neutral couplings involving the massless 
neutrinos are diagonal but flavour-dependent.

It has been shown \cite{brv} that for $n$ generations
the total number of physical parameters describing 
the Yukawa sector is 
$n^2$ angles and $(n-1)^2$ phases. 
Thus, for three families there are four independent CP-violating phases. 
If the charged lepton Yukawas are neglected, it is easy to show, using the
method of ref. \cite{arca}, that the number of phases is
$(n-1)(n-2)/2$. So one CP-violating phase still remains in this limit and 
consequently the associated CP invariant is not suppressed by the charged
lepton masses. In contrast, the CP invariants corresponding to the other 
three phases
are necessarily suppressed by the small differences of charged lepton masses.

\section{Order of Magnitude Estimates}

The observable CP asymmetry results from the interference of
pure CP-vio\-la\-ting phases with  
CP-even phases, equal for particles and antiparticles.
These are the reflection coefficients, which become complex when 
the particle energy is smaller than its mass in the true vacuum.  
Unremovable CP-odd phases appear in the mass matrices due to  
either

a) CP-violating interactions in the thermal loops that correct the dispersion 
relations 
of the particles propagating in the plasma \cite{fs};

b) non-trivial space-time dependence of the scalar vevs
inside the bubble wall (for more than one Higgs field), which
induces space dependent CP-violating phases. These phases cannot be 
rotated away at two adjacent points, $x$ and $x + dx$,
by the same set of unitary transformations, i.e. 
$U_x^{-1} U_{x+dx} \neq 1$ \cite{hn}. 

Whenever mechanism b) is present, it generically will dominate over a), 
since in a) there are suppression factors coming from loops 
$(1/4 \pi)$. 
Mechanism b) is the one that generates the baryon asymmetry in  all
the extensions of the SM proposed in the literature
for electroweak baryogenesis. 
In contrast, in the SM the quark mass matrix has only an overall 
dependence on the Higgs vev and can be 
diagonalized by space-independent unitary matrices; hence  
the CP asymmetry can only be
generated through mechanism a).

In the model considered here we have to distinguish between the  
charged and the neutral sectors. Charged leptons get their masses 
only from the doublet scalar vev, so the situation is completely 
analogous to the SM:
CP-violating phases appear in the thermal corrections to the 
dispersion relations. 
In the neutral sector the situation is different because
the mass matrix has a
non-trivial dependence on both singlet and doublet scalar vevs.
Since generically 
this ratio is not constant within the wall,   
mechanism b) is also present.

The size of the 
leading 
CP asymmetries in the reflection of both charged and
neutral leptons can be estimated by simple power counting arguments. 
To do so, we construct a measure of the CP violation, invariant 
under flavour and phase redefinitions of the lepton fields, i.e. 
under transformations of the type
\bea
\Psi_L & \rightarrow & U \Psi_L  \nonumber\\
e_R & \rightarrow & V e_R  \nonumber\\
\nu_R & \rightarrow & W \nu_R  \\
s_L & \rightarrow & X s_L.  \nonumber
\label{trans}
\eea
One can show that the following expression is invariant under such 
transformations, and vanishes if CP is conserved \cite{brv}: 
\beq
{\rm Im Tr} \left[D^\dagger D m^\dagger m D^\dagger S S^\dagger D \right].
\label{cp2}
\eeq
Notice that this effect cannot be tree level in the reflection 
amplitude, since it involves the couplings 
of both charged 
and neutral leptons.  
Therefore it is typically down by loop factors ($1/4\pi$). 

This invariant is given by
\beq
\delta_{CP}^2  = 
\sum_{a <b} M_a^2 M_b^2 (M_a^2 - M_b^2) \sum_{i < j} (m_i^2 - m_j^2)
{\rm Im} (K_{Hia} K_{Hja}^* K_{Hib}^* K_{Hjb}),
\label{2g}
\eeq
where $M_i$ and $m_a$ are the NHL and charged lepton masses,
respectively.
The natural scale in the problem is of  
the order of the electroweak phase transition temperature,
$T \sim 100$ GeV. Therefore, to obtain a dimensionless quantity  
$\delta_{CP}^2$ should be divided by $T^8$. Typically 
$M_i \sim T$, but the small charged lepton masses give a suppression
of order $(m_\tau/T)^2 \sim 10^{-4}$ at least.
We expect from eq. (\ref{2g}) that the leading CP asymmetry 
in the reflection of charged leptons 
will appear at fourth order in the mixing 
$K_H \sim {\cal O}(D/S)$. 

On the other hand, the leading effect coming from the neutral sector 
appears at tree level.
The leading CP measure, invariant under the transformations of
(\ref{trans}) involving only the
neutral fields, is given by
\beq
{\rm Im Tr} 
\left[D D^\dagger S S^\dagger (D D^\dagger)^2 (S S^\dagger)^2 \right],
\label{cp3}
\eeq
which gives
\beq
\delta_{CP}^3 = 
M_1^2 M_2^2 M_3^2 (M_1^2 - M_2^2) (M_1^2 - M_3^2) (M_2^2 - M_3^2)
%\nonumber \\ &\times
{\rm Im}
\left[(K_H^\dagger K_H)_{12} (K_H^\dagger K_H)_{23} (K_H^\dagger K_H)_{31}
\right]. 
\label{3g}
\eeq
In this case, the asymmetry appears at sixth
order in the mixing, ${\cal O}[(D/S)^6]$.
To obtain a dimensionless quantity, we should divide 
$\delta_{CP}^3$ by $T^{12}$ but there is no suppression in the masses here
since $M_i \sim T$.
There is also an additional contribution of the form (\ref{2g}) due to 
loop corrections, but  it would be
suppressed at least by $(m_\tau/T)^2 \sim 10^{-4}$ and by loop 
factors, which considering the experimental bounds, is  a larger suppression than the extra $(D/S)^2$.

If we assume that the asymptotic value of the ratio of scalar vevs is the 
same as at zero temperature  $v(T)/u(T) = v(0)/u(0)$, there are 
quite strong  experimental bounds on the elements of the
submatrix $K^\dagger_H K_H$. These bounds  depend on the NHL 
mass \cite{eb,nrt}.
For 3 GeV $\le M \le M_Z$  the strongest limits come from LEP  
and they are very stringent
\footnote{These bounds have been obtained 
using inequalities of the form $|{\rm Im} (K_{Hia} K_{Hja}^* K_{Hib}^* K_{Hjb})| \leq \frac{1}{8}[(K^\dagger_H K_H)_{ii}^2 + (K^\dagger_H K_H)_{jj}^2]$.}: 
$|{\rm Im} (K_{Hia} K_{Hja}^* K_{Hib}^* K_{Hjb})| 
\le 3 \times (10^{-9}$--$10^{-7})$.
If $M \ge M_Z$, there are low-energy constraints that 
arise both from the non-observation of lepton 
flavour violation and from universality, as well as limits from the 
invisible width of the $Z$ boson \cite{nrt}. 
The limits are slightly weaker for the mixing of the third 
family with any of the first two, 
\beq
|{\rm Im}(K_{Hia} K_{H3 a}^* K_{Hib}^* K_{H3 b})| 
\le 5 \times 10^{-5} \,(10^{-5}) 
\ \ \ \, i=e,\mu.
\label{eb1}
\eeq
The first number corresponds to the so-called `joint' bounds in 
ref. \cite{nrt}, for which cancellations among the different 
possible fermion mixings are allowed, while the number in brackets
corresponds to the `single' limits, obtained when
the remaining mixing parameters are set to zero.

For the invariant (\ref{cp3}), we get
\beq
{\rm Im}\left[(K_H^\dagger K_H)_{12} (K_H^\dagger K_H)_{23} 
(K_H^\dagger K_H)_{31}
\right ]  \le  10^{-7} \, (3 \times 10^{-8}).
\label{eb2}    
\eeq
Based on the bounds (\ref{eb1})  and taking into 
account the loop factor expected in that case,
plus the further suppression in the charged lepton masses,
typically of order $(m_\tau/M)^2 \sim 10^{-4}$,  
the CP asymmetry in the reflection of charged leptons (\ref{2g}) is expected to be too small to generate a significant baryon asymmetry. 
However, a similar enhancement as the one found in \cite{fs} could imply that 
there is no suppression coming from the lepton mass, in which case the effect
could be important. This is the reason why we decided to do a detailed 
calculation in this case. 

In the case of the reflection of the neutral leptons, the bound (\ref{eb2}),
together with the fact that there is no power suppression in the light masses 
or loop factors, 
implies that the effect could be of roughly the right order of magnitude.

\section{CP Asymmetry in the Neutral Sector}

In this section we will compute the CP asymmetry in the number current of 
$\nu_L$ that get reflected into the unbroken phase. This asymmetry in the rest
frame of the wall is simply
 given by 
\begin{eqnarray}
j_{CP} = j^{trans}_{s^b_L \rightarrow \nu^u_L} + 
j^{trans}_{\nu^b_L \rightarrow \nu^u_L} 
+ j^{ref}_{\nu^u_R \rightarrow \nu^u_L} = \nonumber
\end{eqnarray}
\begin{eqnarray}
\int \frac{d^3 p'}{(2 \pi)^3} \; \sum_{i,j}[|A^t_{{s^i}^b_L \rightarrow {\nu^j}^u_L}(p'_z,
\sqrt{{p'}^2_z + M^2_i})|^2 - |\bar{A}^t_{{s^i}^b_L \rightarrow {\nu^j}^u_L}
(p'_z,\sqrt{{p'}^2_z + M^2_i})|^2] \;\frac{p'_z}{E} f^b_i(p') \nonumber\\
+ \int \frac{d^3 p}{(2 \pi)^3} \; \sum_{i,j}[ | A^t_{{\nu^i}^b_L \rightarrow{\nu^j}^u_L}
(p_z,p_z)|^2 - |\bar{A}^t_{{\nu^i}^b_L \rightarrow {\nu^j}^u_L}(p_z,p_z) |^2 ] 
\;\frac{p_z}{E} f_0^b(p) \nonumber\\
+ \int \frac{d^3 p}{(2 \pi)^3} \; \sum_{i,j}[ |A^r_{{\nu^i}^u_R \rightarrow {\nu^j}^u_L}
(-p_z,p_z) |^2 - |\bar{A}^r_{{\nu^i}^u_R \rightarrow {\nu^j}^u_L}(-p_z,p_z) |^2
 ] \; \frac{p_z}{E} f_0^u(p),
\label{deltacp} 
\end{eqnarray}
with $\sum_{i,j}$ being the sum over flavours and
\begin{eqnarray}
f^u_0(p) = \frac{1}{e^{( |\vec{p}| + v_w\;p_z)/ T} + 1},
\end{eqnarray}
\begin{eqnarray}
f^b_0(p) = \frac{1}{e^{( |\vec{p}| - v_w\; p_z)/ T} + 1}, \;\;\;
f^b_i(p') = \frac{1}{e^{(\sqrt{|\vec{p'}|^2+ M^2_i} - v_w\; p'_z)/T} + 1},
\end{eqnarray}
being the thermal distributions of the different particles in the unbroken (u) 
and broken (b) phases as seen from the rest frame of the wall; $v_w$ is the 
wall velocity, which is estimated to be $v_w \sim 0.1$--$0.4$
in the SM \cite{vw}. 

In the present case, thermal corrections to the propagation are 
negligible. The thermal masses are  
$\leq 0.25 M_i$ for the heavy leptons and thus considerably
smaller than the energies at which the effect will be significant, 
$\omega \geq min\{M_i\}$. Furthermore, the mean free 
path of these weakly interacting particles is expected to be
large compared both to the expected width of the bubble wall and to the 
reflection time of the leptons $\sim M_i^{-1}$; in the scattering
with the wall the neutral leptons will therefore be assumed to be free. 
The transmission and reflection amplitudes will thus be computed at zero
temperature, using LSZ reduction
 formulae in terms of the propagator in the presence of the wall:
\begin{eqnarray}
{\cal A} = \int d^4 x \int d^4 y \;e^{-i q_i x} e^{i q_f y} \;\bar{u}(q_f) 
(i \vec{\partial} - m) S(y, x) ( - i \vec{\partial} - m)  u(q_i) \nonumber\\
= (2 \pi)^3 \delta(q^x_f - q^x_i) \;\delta(q^y_f - q^y_i) \;\delta(E_f - E_i)\;
A(q^z_i,q^z_f),
\label{lsz}
\end{eqnarray}
with
\begin{eqnarray}
S(y, x) \equiv \langle 0| T[ \Psi(y) \bar{\Psi}(x) ] | 0 \rangle .
\end{eqnarray} 
An analogous expression holds for antiparticles. The spinors in formula 
(\ref{lsz}) are on-shell and normalized to unit flux in the $z$ direction, i.e.
\begin{eqnarray}                
\bar{u}\; \gamma_z \;u = 1.
\label{norm}
\end{eqnarray}

Since the potential created by the bubble wall is only dependent on the 
coordinate $z$, momenta in the $x$ and $y$ directions are conserved. 
The transmission and reflection amplitudes only depend on the momenta in the 
$z$ direction and can be computed in a much simpler way by first boosting to
a  frame where $q_x , q_y = 0$. With the proper normalization chosen for the 
spinors (\ref{norm}), the amplitude in the boosted frame is simply given
by (\ref{lsz}), with the propagator and incoming and outcoming momenta 
substituted by the boosted ones. 

We can further simplify the expression for $j_{CP}$ by using CPT symmetry 
and unitarity constraints, which imply 
\begin{eqnarray}\sum_i | A^t_{{\nu^i}^b_L \rightarrow {\nu^j}^u_L} |^2  
= 1 - \sum_i | A^r_{{\nu^i}^u_R \rightarrow {\nu^j}^u_L} |^2 
- \sum_i | A^t_{{s^i}^b_L \rightarrow {\nu^j}^u_L} |,
\label{shit}
\end{eqnarray} 
and substituting (\ref{shit}) in eq. (\ref{deltacp}):
\begin{eqnarray}
j_{CP} = \int \frac{d^3 p}{(2 \pi)^3} \;\sum_{i,j}( | A^r_{{\nu^i}^u_R \rightarrow {\nu^j}^u_L} |^2 - | \bar{A}^r_{{\nu^i}^u_R \rightarrow {\nu^j}^u_L} |^2 ) \;
 \frac{p_z}{E} ( f_0^u(p) - f_0^b(p) )  \nonumber\\
+ \int \frac{d^3 p}{(2 \pi)^3} \;\sum_{i,j} (| A^t_{{s^i}^b_L \rightarrow {{\nu^j}}^u_L} |^2 - | \bar{A^t}_{{s^i}^b_L \rightarrow {{\nu^j}}^u_L} |^2  ) \;\frac{p_z}{E} ( f_i^b(p') - f_0^b(p) ).
\end{eqnarray}
Finally, by expanding the Fermi distributions for small wall velocities: 
\begin{eqnarray}
j_{CP} = \frac{v_w}{T} \int \frac{d^3 p}{(2 \pi)^3} \left\{ - 2 \;\sum_{i,j}
( | A^r_{{\nu^i}^u_R \rightarrow {\nu^j}^u_L} |^2 - 
| \bar{A}^r_{{\nu^i}^u_R \rightarrow {\nu^j}^u_L} |^2  ) \right.\nonumber\\
\left. + \sum_{i,j} (\frac{\sqrt{p^2_z-M^2_i}-p_z}{p_z} ) (| A^t_{{s^i}^b_L 
\rightarrow {{\nu^j}}^u_L} |^2 - 
| \bar{A}^t_{{s^i}^b_L \rightarrow {{\nu^j}}^u_L} |^2 )  \right\} 
\frac{{p^2_z}}{E} f_F(p) [1 - f_F(p) ],  
\label{jcp}
\end{eqnarray}
where $f_F(p) = 1/ ( e^{ |\vec{p}|/T} + 1)$ is the unboosted Fermi distribution.

In order to compute the amplitudes in eq. (\ref{lsz}) we would need the exact
propagator in the presence of the wall. The potential created by the wall in 
the weak basis ($\nu_L$, $s_L$, $\nu_R$) is
\begin{eqnarray}
M(z) = 
\left(
\ba{ccc}
0 & 0 & V(z) K_H M\\
0 & 0 & U(z) K_{SH} M\\
V(z) M K^{\dagger}_H & U(z) M K^{\dagger}_{HS} & 0
\ea
\right),
\label{mwb}
\end{eqnarray}
where $V(z) \equiv v(z)/v$ is the ratio of the vevs of 
the doublet scalar $H$ in the wall and the asymptotic vev in the broken phase;
 $U(z)$ is 
the ratio corresponding to the singlet $\sigma$ field and $M = (M_i)$ is the diagonal mass matrix of the Dirac neutrinos. 

The simplest approach would be to do perturbation theory  in $M(z)$ 
\cite{hs,hn}, which is effectively an expansion in $M(z)/\omega$. 
Although this approximation makes the calculation much simpler, 
it is not justified since the region of interest is always $\omega \sim M_i$. 
Instead, we will perturb in the mixing, that is in $K_H = {\cal O}(D/S)$ and 
$K_{HS} - 1 = {\cal O}[(D/S)^2]$. We can write the mass matrix as
\begin{eqnarray}
M(z) = M_0(z) + \delta M(z),
\end{eqnarray}
with
\begin{eqnarray}
M_0(z) = 
\left(
\ba{ccc}
0 & 0 & 0\\
0 & 0 & U(z) M\\
0 & U(z) M & 0
\ea
\right),\;\;\; \delta M(z) = M(z) - M_0(z).
\label{mwbexp}
\end{eqnarray}

Our strategy will be to solve the scattering problem with potential $M_0(z)$  
exactly and perturb only in $\delta M(z)$. This can be done for several 
forms of $U(z)$. In order to simplify the problem we consider here  the thin 
wall approximation for the singlet field i.e.,  $U(z) = \theta(z)$. The 
result will be more important for a singlet width as different from the 
doublet width as possible, so we will keep the singlet 
width to its minimum value and vary the doublet one. There is no reason 
to expect that any other choice would give very different results. 
The perturbation in 
$\delta M$ then gives
\begin{eqnarray}
S(x_2, x_1)= \int \prod_i d z_i \;S^{(0)}(x_2,z_1) \delta M(z_1) 
S^{(0)}(z_1,z_2) \delta M(z_2) \ldots S^{(0)}(z_i,x_1), 
\end{eqnarray}
where the integration is done over all $z_i (-\infty, \infty)$ and $S^{(0)}$ 
is the exact propagator in the potential $M_0(z)$. It is a matrix with spin 
structure
\begin{eqnarray}
S^{(0)} = 
\left(
\ba{ccc}
S^{(0)}_{\nu_L} & 0 & 0\\
0 & S^{(0)}_{LL}& S^{(0)}_{LR}\\
0 & S^{(0)}_{RL} & S^{(0)}_{RR}
\ea
\right).
\label{prop}
\end{eqnarray}
We can then use the zero temperature propagator in the presence of a thin 
wall that has been computed in \cite{b1}. The formulae are given 
in appendix A. 
                                                         
We approximate the doublet field wall profile, $V(z)$, as 
\beq
V(z) =
\left \{
\ba{cc}
0 &  z < 0   \\
\delta_H^{-1} z & 0 < z <  \delta_H  \\ 
1 & z > \delta_H
\ea  \right. 
\eeq
so that the wall thickness is parametrized by $\delta_H$.
We expect that this simple form is enough to give a reasonable
estimate of the CP asymmetry.                         

The calculation is straightforward. In the case of three families, 
the result turns out to be non-zero at sixth order in $(D/S)$ 
as expected 
from the invariant (\ref{cp3}). The contribution coming from the reflection 
is 
\begin{eqnarray}
\sum_{i,j}  (| A^r_{{\nu^i}^u_R \rightarrow {\nu^j}^u_L} |^2 
- {\rm antiparticles} )  =  \; J_{ijk} \; F^r(M_i,M_j,M_k),
\label{r}
\end{eqnarray}
and the transmission one 
\begin{eqnarray}
\sum_{i,j} ( | A^t_{{s^i}^b_L \rightarrow {\nu^j}^u_L} |^2 - {\rm antiparticles}) =
\; J_{ijk} \; F^t(M_i,M_j,M_k),
\label{tr}
\end{eqnarray}
where
\bea
J_{ijk} &=& {\rm Im}[(K^{\dagger}_H K_H)_{ij} (K^{\dagger}_H K_H)_{ki} 
(K^{\dagger}_H K_H)_{jk}]\nonumber\\
F^r(M_i,M_j,M_k) &=& 2 M^2_i M^2_j M^2_k \{ Im[ A^*_{ik} A_{ij}] 
- 2 {\rm Im}[  I^{i*}_{1a} \frac{I_{5a}^{ijk}}{4 p_j p_k} )] \}\nonumber\\
F^t(M_i,M_j,M_k) &=& - M^2_i M^2_j M^2_k  
\frac{|M_i + \omega - p_i |^2}{2 p_i \; (\omega+M_i)} 
{\rm Im}
\left \{ \frac {I^{ij*}_{3b}}{2 p_j^*} \frac {I^{ik}_{3b}}{2 p_k}
+ 2  I^{i*}_{1b} \frac{I_{5b}^{ijk}}{4 p_j p_k} ) \right \},
\eea
and
\beq
A_{ij} \equiv \frac{1}{2 p_j} (I^{ij}_{3a} - i I^{ij}_{2a}) \;\;\;
p_i \equiv \sqrt{\omega^2 - M^2_i}.
\eeq
The integrals $I_{1a},\ldots, I_{5a}$ are defined in appendix B.

It can be easily checked that whenever two masses are degenerate 
the result vanishes.
We have also checked that in the thin wall approximation,
i.e. $\delta_H \rightarrow 0$, the effect disappears, as it should happen
since in this limit both the singlet and doublet wall profiles
become the same.

\begin{figure} 
\begin{center}
\psfig{figure=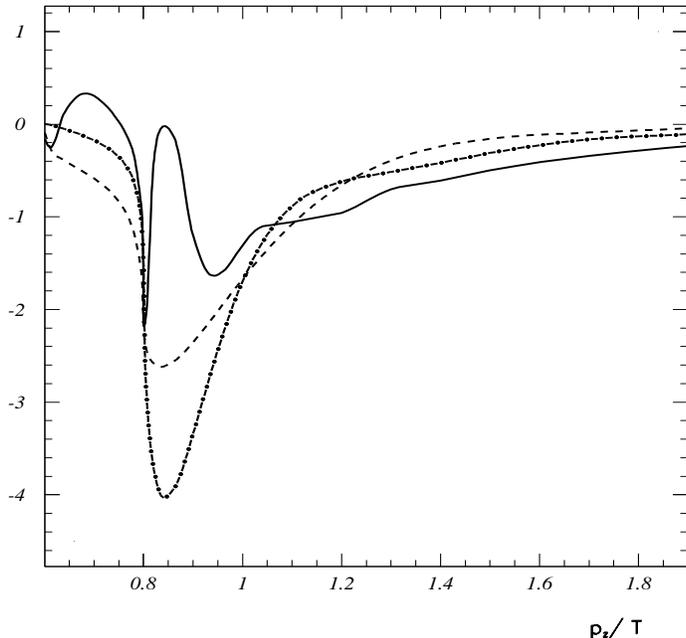,rwidth=2.5in,rheight=2.5in,width=4in,height=5in,bbllx=60pt,bblly=-30pt,bburx=500pt,bbury=600pt}
\end{center} 
\vspace*{5mm}
\caption{$\epsilon_{ijk} F^t(M_i,M_j,M_k)$ as a function of $p_z$ for 
NHL masses $M_i = (.6, .8, 2.)$ and 
$\delta_H^{-1} =.1$ (solid),  $\delta_H^{-1} =.2$ (dashed-dotted) and
$\delta_H^{-1} =.3$ (dashed). All in units of the temperature.}
\end{figure} 

\begin{figure} 
\begin{center}
\psfig{figure=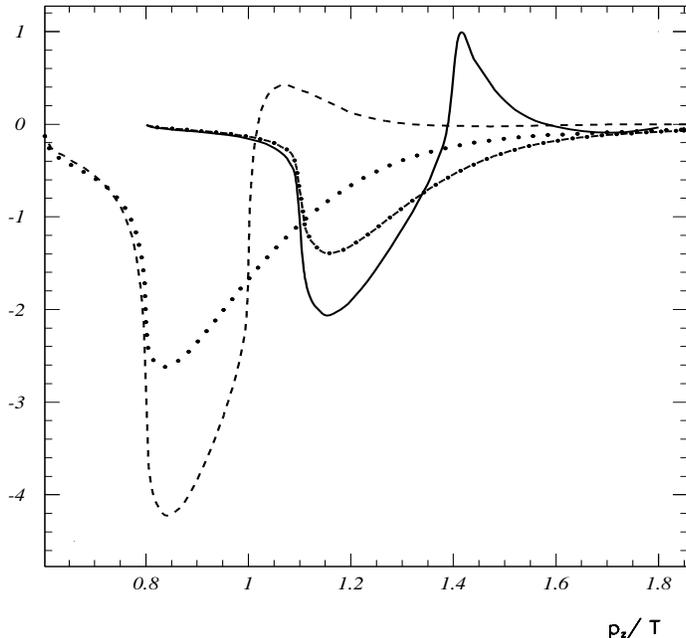,rwidth=2.5in,rheight=2.5in,width=4in,height=5in,bbllx=60pt,bblly=-30pt,bburx=500pt,bbury=600pt}
\end{center} 
\vspace*{5mm}
\caption{$\epsilon_{ijk} F^t(M_i,M_j,M_k)$ as a function of $p_z$
for $\delta_H^{-1}= .3$ and NHL masses: 
i) (.8, 1.1, 1.4) solid line, ii) (.8, 1.1, 2.) dashed-dotted,
iii) (.6, .8, 1.) dashed and iv) (.6, .8, 2.) dotted. All in units of the 
temperature.} 
\end{figure}

As we saw in the previous section, 
for three generations
there is only one CP-violating phase and
$J_{ijk} = J_{123}\; \epsilon_{ijk}$. In this case, the phase
$J_{123}$ factorizes in the asymmetries of eqs. (\ref{r}) and
(\ref{tr}). We find that the contribution from the transmission
amplitude is the dominant one, while the reflection amounts 
to a small correction.
In figs. 1 and 2 we plot $\epsilon_{ijk} F^t(M_i,M_j,M_k)$ as
a function of $p_z$, for different values of the NHL masses and 
$\delta_H$.
We will consider masses of order 
$\sim T$, because for heavier neutral leptons the
current will be strongly 
suppressed by the Fermi distribution in eq. (\ref{jcp}).

Figure 1 shows the dependence on $\delta_H$ for a fixed NHL spectrum. 
We find that the effect is more important when the 
mass differences of at least two NHLs are  $(M_i-M_j) \leq \delta_H^{-1}$.
When all mass differences are larger than $\delta_H^{-1}$,
 the asymmetry oscillates rapidly (we expect the oscillation 
period to be related to $\delta_H$) and the integrated 
result is suppressed. Also we observe that the effect is smaller 
for smaller $\delta_H$, in agreement with the fact that 
it vanishes in the limit $\delta_H \rightarrow 0$. Thus we 
expect that the largest effect will occur
for $\delta_H$, satisfying $(M_i-M_j) \leq \delta_H^{-1}$, as large as possible.

In fig. 2 we fix $\delta_H$ and keep $M_2-M_1 \leq \delta_H^{-1}$, 
while varying the masses $M_1$ and  $M_3$. The peak appears 
after the second threshold (i.e. $p_z > M_2$); thus, as $M_2$ changes, the 
position of the peak moves accordingly. Nevertheless, 
the integrated result is not very sensitive to the particular 
values of the NHL masses, provided the relation 
$M_2-M_1 \leq \delta_H^{-1}$ is satisfied. 

The current $j_{CP}$ we obtain for generic values of the masses,
with the only restriction that $M_i \sim T$ and $M_i - M_j \sim \delta_H^{-1}$
for at least two NHL, is typically $(0.5$--$1) \times 10^{-2} J_{123} \;v_w$.

\section{CP Asymmetry in the Charged Sector}

In this section we compute the CP asymmetry in the flux of
charged left-handed leptons, $l_L$, reflected in the unbroken phase.
The charged lepton mass matrix has just an overall dependence on the 
doublet scalar vev (as occurs in the SM); 
therefore only mechanism a) as defined in section 3 is present in this case. 

The calculation of the charged lepton CP asymmetry 
is completely analogous to the computation done for quarks
in the SM \cite{fs,b2,hs}, and we refer the reader to these
works for further details. Contrary to the case of the NHL, the one-loop thermal corrections are much larger than the tree-level masses 
of the charged leptons.
The resummation of the thermal self-energies considerably modifies the
dispersion relations and
 the correct asymptotic states are now quasi-particles.

Following the notation of ref. \cite{w}, the thermal one-loop 
contribution to the charged lepton self-energy 
in the broken phase can be written as 
\beq
{\rm Re} (\Sigma(k)) = -a \not\!k - b \not\!u,
\label{se}
\eeq
where $a$ and $b$ are matrices in flavour space, $u$ is the 
four-velocity of the plasma and $k=(\omega,{\bf k})$ is the 
external momentum. We have neglected the contribution 
proportional to the masses of the charged leptons.
In the plasma rest frame and the mass basis,
\beq
{\rm Re} (\Sigma(\omega,{\bf k})) \gamma_0 =
-h(\omega,{\bf k}) - a(\omega,{\bf k}) {\bf \alpha} \cdot {\bf k},
\eeq
where $h(\omega,{\bf k})= a(\omega,{\bf k}) \omega + b(\omega,{\bf k})$
and is given by
\beq
h_{ji} = - f_\gamma H(m_i,0) \delta_{ji}
-\frac{g^2} {2} 
\left \{ [f_Z H(m_i, M_Z) + f_H H(m_i, M_H)] \delta_{ji} 
+ \sum_\alpha f_{W,\alpha} H(M_\alpha, M_W) \right\},
\label{h}
\eeq
with
\bea
f_\gamma &=& Q_i^2 g^2 s_W^2 (L+R)  \\
f_{W,\alpha} &=& \left ( 1 + {\lambda_\alpha^2 \over 2} \right) L
K_{i \alpha}^*  K_{j \alpha} + {\lambda_i \lambda_j \over 2} R
K_{i \alpha}^*  K_{j \alpha}  \\
f_Z &=& {1 \over 2}  \left [ {4 \over c_W^2} (T_i^3 - Q_i s_W^2)^2 
+ {\lambda_i^2 \over 2} \right ] L + {1 \over 2}  \left [ 
{4 \over c_W^2} (Q_i s_W^2)^2 + {\lambda_i^2 \over 2} \right ] R \\
f_H &=& {\lambda_i^2 \over 4} (L+R)
\eea
where $L,R$ are the chiral projectors,
$\lambda_i = m_i/M_W$, $m_i$ are the masses of the external
flavours, and $M_\alpha$ are the masses of the neutral 
leptons inside the loop.
The function $H(M_F,M_B)$ can be found in \cite{b2}.

The dispersion relations of the quasi-particles are then given by
\beq
\not\!k - {\rm Re} (\Sigma(k)) = 0.
\eeq
Since these no longer are  Lorentz invariant, it is not possible to simplify the calculation of
the reflection amplitudes by boostings them to the frame where 
$k_x, k_y =0$, as we did in section 4. The realistic 
computation in three dimensions then becomes very involved. 
However, since our main interest is to study whether 
the enhancement found in \cite{fs} is present, and this can already
be seen in the one-dimensional problem, 
we restrict our discussion to this  
simpler case.

Our objective is to compute 
the number current of $l_L$ reflected on the wall, which for
  small wall velocities and using unitarity and CPT, is
given by
\beq
j_{CP} = -2 \; \frac{v_{w}}{T} \int {d \omega  \over 2 \pi}
 \omega \; v_g^2 \; f_F(w) [1 - f_F(w) ] \Delta_{CP}(\omega), 
\label{jz} 
\eeq
where $v_g \equiv \frac {\partial \omega} {\partial k} = 1/3$ is 
the group velocity, 
\beq
\Delta_{CP}(\omega) =
\sum_{i,j}
( | A^r_{{l^i}^u_R \rightarrow {l^j}^u_L} |^2 - 
| \bar{A}^r_{{l^i}^u_R \rightarrow {l^j}^u_L} |^2  ),          
\eeq
with $A^r$ the reflection amplitudes  on the wall,
and $f_F(\omega)$ is the unboosted Fermi distribution of quasi-particles 
in the plasma rest frame:
\beq
f_F(\omega) = 1/ ( e^{ \omega/T} + 1).
\eeq

To lowest order in the wall velocity,
the scattering problem can be approximately solved  in the rest frame of the wall, neglecting the corrections to the dispersion relations of the 
quasi-particles due to the small boost (which are proportional to 
$v_w$ and are negligible at lowest order).
In this frame, to leading order in $T$ and neglecting the 
flavour-non-diagonal
corrections in (\ref{h}), the quasi-particles propagate 
according to the effective Hamiltonian
\beq
H^0_{eff} = \theta(-z) 
\left(
\ba{cc}
-{i \over 3}  \sigma_z \partial_z + \w^u_R & 
0 \\
0 &
{i \over 3}  \sigma_z \partial_z + \w^u_L
\ea
\right)
+ \theta(z) \left(
\ba{cc}
-{i \over 3}  \sigma_z \partial_z + \w^b_R & 
{m \over 2}  \\
{m \over 2}  &
{i \over 3}  \sigma_z \partial_z + \w^b_L
\ea
\right),
\label{ham}
\eeq                                                
where $\w^{u(b)}_{L,R}$, satisfying
\beq
\w^{u(b)}_{L,R} + \bar{h}^{u(b)}_{L,R}(\w^{u(b)}_{L,R}, 0) = 0,
\eeq
are the thermal masses in the unbroken and broken phases respectively (the 
functions $\bar{h}$ contain only the leading $T$ flavour-diagonal corrections in
(\ref{h})). 
This effective Hamiltonian is only valid for
 low momentum compared to the thermal masses $\w^{u(b)}_{L,R}$. Since 
the reflection of quasi-particles on the wall will occur for $k_z \leq m << \omega^{u(b)}_{L,R}$, this approximation is justified. 

In order to obtain a non-vanishing CP asymmetry, both the subleading corrections in $T$ (which 
introduce the dependence on the NHL masses) and 
the flavour-non-diagonal terms (which contain the mixings) in (\ref{h}) are needed. 
After including these corrections, we get the following effective Hamiltonian:
\beq
H_{eff} = H^0_{eff} + \frac{1}{2} 
\left(
\ba{cc}
\theta(-z) \delta h^u_R(\omega^0, 0)+ \theta(z) \delta h^b_R(\omega^0, 0) & 
0 \\
0 &
\theta(-z) \delta h^u_L(\omega^0, 0)+ \theta(z) \delta h^b_L(\omega^0, 0)
\ea
\right)
\label{hamt}
\eeq                                                
where $\delta h \equiv h - \bar{h}$,
 contain the subleading effects in $T$ and the  flavour-non-diagonal
electroweak corrections. 
The reflection amplitudes of quasi-particles on the wall can then
be obtained by first
solving  for eigenstates of the unperturbed Hamiltonian (\ref{ham}),
which are superpositions 
of  incoming,  reflected and  transmitted plane waves, and then 
perturbing  in the extra terms of (\ref{hamt}).

Up to now, we have neglected the imaginary part of the one-loop self-energy 
(\ref{se}). This contribution
is proportional to the damping rate of the quasi-particles, i.e. their inverse
 lifetime.
There is no calculation of the damping rate $\gamma$ 
of leptons in the SM, but from the result for pure $SU(2)$ at 
zero momentum \cite{dr}, we can estimate
$\gamma \sim \aw  T$, i.e.  $\gamma \sim 1$ GeV at $T=100$ GeV.
In refs. \cite{b2,hs} it was shown that the damping effects for quarks in 
the SM lead to a sizeable suppression of the CP asymmetry, because 
the lifetime $\sim 1/ (2 \gamma)$ of the quasi-quarks in the plasma, was 
much smaller than their reflection time on the wall $\sim 1/m$ (for the down 
quarks, which gave the leading contribution). In the present case, we expect 
that the main effect will come from the reflection of the $\tau$ lepton and,
according to the previous rough estimate, the lifetime of the quasi-tau would
be of the same order of magnitude as its reflection time. In this situation, 
it is not clear whether the damping will have an important effect or not.  
We will first compute the asymmetry neglecting the damping completely and 
at the end of this section we will estimate its effect. As we will see, it
 leads to a suppression that varies rapidly with the exact value 
of $\gamma$ around  the region $\gamma \sim m_{\tau}$.

Similarly to the SM case, we find that the first effect in the 
asymmetry appears at ${\cal O} (\aw^2)$.
Defining $r_0 \equiv A^r_0$ (the unperturbed reflection amplitude), we get
\bea
\Delta_{CP}^{(2)}(\omega) &\sim&
\sum_{i,j} {\rm Im} [(\delta h^b_L)_{ji} (\delta h^b_R)_{ij}] 
\nonumber \\
&\times&  {\rm Im} \left\{ r^{0*}_{ii} \left[
{r^0_{jj} \over  |d_{ij}|^2} +
\frac{m_j [(r^0_{ii})^2-(r^0_{jj})^2]}{2 d_{ii} d_{ij} d_{ji}} +
{r^0_{jj} \over d_{ii}}({1 \over  d_{ij}} + {1 \over d_{ji}})
\right] \right\},
\label{res}
\eea
where $\delta h^b$ are  flavour-dependent and $d_{ij} \equiv \omega^i_L + \omega^j_R - 2 \omega  +
\frac{m_i r^0_{ii}}{2} + \frac{m_j r^0_{jj}}{2}$.
From eq. (\ref{res}) we see that, just as in the SM, 
the effect comes from the interference 
of the ${\cal O} (\aw)$ terms in $\delta h^b_L$ and 
$\delta h^b_R$, and there is no effect at leading order 
in $T$, because at this order $\delta h^b_R=0$.

Substituting the expressions for $\delta h^b_{L,R}$, 
$\Delta_{CP}^{(2)}(\omega)$ can be written as 
\beq
\cp^{(2)}(\w) = \aw^2   \sum_{i,j} \sum_{a<b} 
{\rm Im} (K_{Hia} K_{Hja}^* K_{Hib}^* K_{Hjb}) 
f(m_i,m_j) F(M_a,M_b), 
\label{cp1}
\eeq
where 
\beq
f(m_i,m_j)= \lambda_i \lambda_j 
{\rm Im} \left\{ r^{0*}_{ii} \left[
{r^0_{jj} \over  |d_{ij}|^2} +
\frac{m_j [(r^0_{ii})^2-(r^0_{jj})^2]}{2 d_{ii} d_{ij} d_{ji}} +
{r^0_{jj} \over d_{ii}}\left({1 \over  d_{ij}} + {1 \over d_{ji}}\right)
\right] \right\}
\eeq
and
\bea
F(M_a,M_b) &=& 
\left[ (2 + \lambda_a^2) I(M_a) - 2 I(0) \right]_{\w_L}
[I(M_b)-I(0)]_{\w_R} 
\nonumber \\
&-& \left[(2 +\lambda_b^2)I(M_b) - 2 I(0) \right]_{\w_L}
[(I(M_a)-I(0)]_{\w_R} .
\eea 
$I(M_a) = \frac{\pi}{2} H(M_a,M_W)$ and the subscript $\omega_{L,R}$ 
indicates at which value of $\omega$ the function $H$ is 
evaluated. 
Equation (\ref{cp1}) shows explicitly the GIM cancellation for both 
external and internal lepton masses.

Of all the terms in eq. (\ref{cp1}), the one corresponding to the pair
of external flavours ($\mu$, $\tau$) gives the largest contribution,  
because $f(m_\mu,m_\tau)$ is a few orders of magnitude larger than for 
the other combinations, while the 
experimental bounds on the mixings are of the same order. We restrict to this leading 
term for which the `joint' bound described in section 3 is 
\beq
|{\rm Im}(K_{H2 a} K_{H3 a}^* K_{H2 b}^* K_{H3 b})| 
\le 5 \times 10^{-5},
\label{mi}
\eeq
independently of the flavour of the heavy leptons $a,b$. Thus, if we assume
that all the mixings (\ref{mi}) are of the same order of 
magnitude, the size of 
the various terms in the sum over the heavy flavours depends only  
on the function $F(M_a, M_b)$. We consider just one of these terms as a 
prototype, i.e.
\beq
\Delta_{CP}^{23ab} = \aw^2 {\rm Im} (K_{H2a} K_{H3a}^* K_{H2b}^* K_{H3b}) 
f(m_2,m_3) F(M_a,M_b),
\label{sim}
\eeq
where there is no sum over $a,b$.
 Since we will allow the two heavy masses $M_a, M_b$ to vary arbitrarily, the 
largest value obtained for the integrated asymmetry, considering only 
(\ref{sim}), is also  an upper bound for the other terms. Thus 
  the final result will be at most three times larger, if the terms add 
coherently.

\begin{figure} 
\begin{center}
\psfig{figure=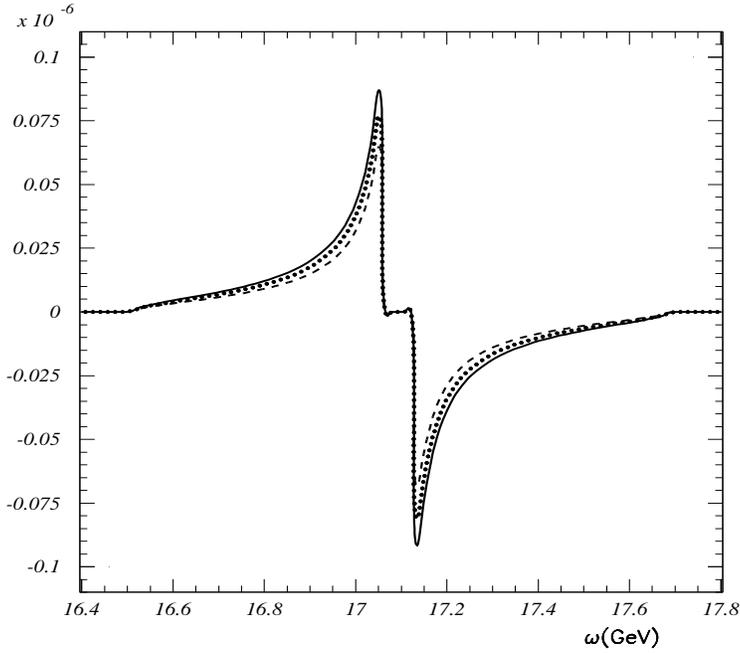,rwidth=2.5in,rheight=2.5in,width=4in,height=5in,bbllx=60pt,bblly=-30pt,bburx=500pt,bbury=600pt}
\end{center} 
\vspace*{5mm}
\caption{CP asymmetry, $\Delta_{CP}^{23ab}(\w)$, for zero damping rate and 
NHL masses $M_a=80$ GeV and
$M_b=$ 110 GeV (dashed), 140 GeV (solid) and 200 GeV (dotted).} 
\end{figure}

In fig. 3 we show the contribution to the CP asymmetry, $\Delta^{23ab}_{CP}(\omega)$.
We have taken the following values for the masses at the phase 
transition temperature ($T \sim 100$ GeV):
$m_{\mu} = 69$ MeV, $m_{\tau} = 1.176$ GeV, $M_W =50$ GeV, 
and the weak coupling is $\aw = 0.035$. 
We have fixed the mass of one NHL to $M_a=80$ GeV, and
we plot the result for different values of the other NHL mass.

The peaks are situated in regions where the $\tau$ lepton reflects 
completely, while the $\mu$ does not. The amplitude of the peaks 
is larger than one would expect from naive power counting, implying
that the suppression in the charged lepton masses is not at work, as found
in \cite{fs}. However, in contrast to what was obtained in the SM,  the two peaks tend to cancel each other,
 and there is a big suppression in the integrated result, since
the Fermi factors in eq. (\ref{jz})
are approximately constant. 
For $M_b = 140$ GeV the contribution of this term to the integrated result in (\ref{jz}) is
$\sim  10^{-12} v_{w}$ (which turns out to be of the same 
order as the naive estimate). Whether the peaks come with 
equal or opposite signs seems to be
very dependent on the relative position of the thermal masses of the different
flavours. In this case the thermal masses are almost flavour-independent, while in  the down sector of the SM  there is a big shift in 
the third family thermal masses compared to the other two, due to the 
top Yukawa.
  This is why there is no such cancellation in that case.  The conclusion is
that the enhancement found in \cite{fs} is rather model-dependent and it seems to
require large flavour-dependent thermal corrections.

\begin{figure} 
\begin{center}
\psfig{figure=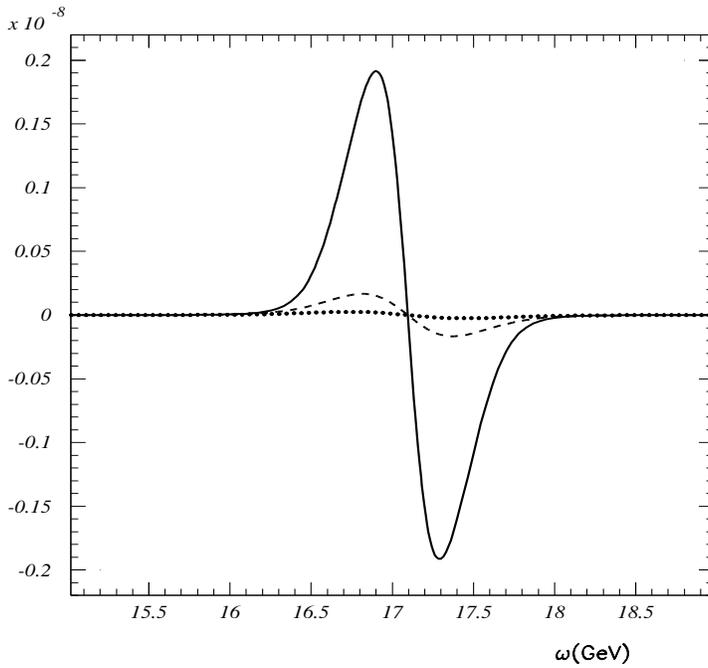,rwidth=2.5in,rheight=2.5in,width=4in,height=5in,bbllx=60pt,bblly=-30pt,bburx=500pt,bbury=600pt}
\end{center} 
\vspace*{5mm}
\caption{$\cp^{23ab}(\w)$ for $M_a= 80$ GeV, $M_b=140$ GeV and 
damping rate $\gamma=.5$ GeV (solid), 1. GeV (dashed) and 1.5 GeV (dotted).} 
\end{figure}

Finally, we want to estimate the effect of the damping rate 
which, as discussed before, is not negligible compared to the reflection
time of the $\tau$ lepton. 
As shown in refs. \cite{b2,hs}, the decoherence effects induced by the
damping rate can be taken into account by including the imaginary 
part of the self-energy in the effective Hamiltonian (\ref{ham})
and solving it for spatially damped waves.               
Since the exact value of the damping rate is not known, we have
computed $\Delta^{23ab}_{CP}(\omega)$ for different values of $\gamma$,
namely $\gamma=$ 0.5, 1., and  1.5 GeV.
The result is shown in fig. 4. As is clear from the curves, the 
suppression due to the damping increases rapidly when
$\gamma \sim m_{\tau}$. Without a precise  determination of 
this quantity,
it is thus impossible to estimate the actual suppression, although it is 
clear that neglecting this effect is not justified. 
 
To summarize, we have found that the CP asymmetry in the reflection of the 
charged leptons is
at most of ${\cal O}(10^{-12})$, even neglecting decoherence effects due to 
interactions in the plasma.

\section{Baryon Asymmetry}

In this section, we calculate the 
baryon asymmetry induced by the CP asymmetries computed in the previous 
sections. This is a very difficult problem since a microscopic treatment is no 
longer possible and we have to match somehow the microscopic result 
of $j_{CP}$ with the thermodynamic treatment of transport of the
chiral lepton number generated at the wall. Strictly speaking, the two 
problems, reflection and transport, are completely coupled and should be solved 
at the same time. This however implies treating a many-body non-equilibrium 
quantum system, and some approximations are necessary. 

We will consider here only the effect obtained from the reflection of the 
neutral leptons, since the CP asymmetry  of the charged leptons computed in the 
last section is far too 
small. In the neutral sector, we have completely neglected the thermal incoherence 
effects 
in the reflection. This has been shown to be a very bad approximation 
when the damping rate is comparable to both the height and/or width of the
wall \cite{b2}\cite{hs}\cite{hn}. However, this is not the case here as we
discussed in section 3.
We believe that, because of this,  reflection can be treated independently of 
transport \footnote{If the damping rate is not small compared
to other scales in the problem, we do not think one can separate the problems
of reflection and transport, and a detailed calculation is much more 
complicated.}.

The picture is then that near the wall in the unbroken phase, a 
local density of $\nu_L$ lepton number is generated due to reflection. This 
local density generates a diffusion current in the plasma and decays due 
to sphaleron processes that take place in the unbroken phase as we go away 
from the wall, generating a baryon number density. This picture is 
only consistent if the reflected particles have enough time
to diffuse before the wall catches up. 
This  will be true for small  velocities of the wall. 
In this case also the incoming 
flux of particles in the calculation of $j_{CP}$ can be taken to be the 
thermal one, as we assumed in the previous sections.

The diffusion equations in the wall rest frame read
\bea
\left(
\ba{c}
\partial_t \;n_B\\
\partial_t \;n_L
\ea \;
\right) = \;
\left(
\ba{cc}
D_B \partial^2_z - v_{w} \partial_z - 3 \Gamma \theta(-z)
& -\Gamma \theta(-z)\\
-3 \Gamma \theta(-z) 
& D_L \partial^2_z - v_{w} \partial_z - \Gamma \theta(-z)
\ea
\right)
\left(
\ba{c}
n_B\\
n_L
\ea 
\right)
\label{diffusion}
\eea
where $\Gamma \equiv 9 \frac{\Gamma_{WS}}{T^3}$ and 
$\Gamma_{WS} = \kappa (\aw T)^4$ is the weak sphaleron rate 
with $\kappa$ a coefficient of ${\cal O}(1)$ \cite{sph}
\footnote{However, there is a recent claim that damping effects in 
the plasma suppress the sphaleron rate to ${\cal O} (\aw^5 T^4)$
\cite{asy}.}. 
We have made the further approximation that the sphaleron
rate in the broken phase is zero. This drastic approximation can only be 
justified if the phase transition is strongly first order; 
$v_w$ is the velocity of the wall.
The constants $D_{L,B}$ are the diffusion coefficients for leptons and 
quarks respectively. Since quarks suffer strong interactions, it is clear 
that
\beq
D_B \ll D_L.
\label{ineq1}
\eeq
We take the values of the diffusion constants estimated in the SM in 
ref. \cite{jpt}, namely  $D_L \sim 110/T$ and $D_B \sim 6/T$.
These estimates are obtained from the elastic scattering, $t$-channel 
vector boson exchange diagrams, which are expected to dominate the 
scattering process. Yukawa interactions are neglected.
We have not included here any other possible $\nu_L$ number decay process 
than the sphalerons. Other decays through Higgs interactions are
obviously possible, but we find that their rate is smaller than the 
sphaleron rate in the unbroken phase, so we can safely neglect them. 

We look for stationary solutions, i.e. $\;\partial_t n_{B,L} = 0$. 
In order to solve the equations for $n_{L,B}$ 
(\ref{diffusion}), we need to impose boundary conditions on the 
densities and their derivatives (diffusion currents). 
We will require that $n_{L,B}(-\infty) = 0$, since there is no 
asymmetry in the incoming thermalized flux seen by the wall. At $z \rightarrow 
\infty$, we require that the solutions be constants. These would be precisely 
the values of $L$ and $B$ in the broken phase that will survive the phase 
transition. At the interphase $z = 0$, we impose continuity of the 
diffusion current, 
\bea
D_{L,B} \partial_z n_{L,B} - v_{w} n_{L,B} \; |^{+}_{-} \; =\; 0
\label{bc1}
\eea  
and the existence of the reflected flux is taken into account in a 
constraint on the lepton density in the unbroken phase near the wall,
\bea
n_L |_{z=0^{-}} \; = \; n^0
\label{bc2}
\eea 
where $n^0 = j_{CP}/\langle v_i \rangle$ and $\langle v_i \rangle$ is the average velocity of the 
particles in the reflected flux that we define as,
\bea
\langle v_i \rangle \equiv \frac{\int \frac{d^3 p}{(2 \pi)^3} 
\;\sum_{i,j} J_{ijk}
\{\; -2 F^r(M_i,M_j,M_k) + (\frac{\sqrt{p^2_z-M^2_i}-p_z}{p_z} ) F^t(M_i,M_j,M_k) \;\} 
\frac{{p^3_z}}{E^2} f_F(p) [1 - f_F(p) ]}{\int \frac{d^3 p}{(2 \pi)^3} 
\;\sum_{i,j} J_{ijk}
\{\; -2 F^r(M_i,M_j,M_k) + (\frac{\sqrt{p^2_z-M^2_i}-p_z}{p_z} ) F^t(M_i,M_j,M_k) \;\} 
\frac{{p^2_z}}{E} f_F(p) [1 - f_F(p) ]}\nonumber
\eea

It is straightforward to obtain the most general solutions \cite{fs} of 
(\ref{diffusion}) in the approximation,
\bea 
\frac{3 D_{B,L} \Gamma}{v^2_{w}} \; \ll 1,
\label{ineq2}
\eea
which  is expected to be of ${\cal O}(10^{-1})$. The solution is 
\begin{eqnarray}
\ba{cc}
n_B = C_1 a_{11} e^{k_1 z} + C_2 a_{21} e^{k_2 z} & \nonumber\\
n_L = C_1 a_{12} e^{k_1 z} + C_2 a_{22} e^{k_2 z} & z < 0\nonumber\\
n_B = B &  \nonumber\\
n_L = L & z > 0
\ea
\end{eqnarray}
with
\bea
k_1 \equiv \frac{v_{w}}{D_B} \left( 1+ \frac{3 \Gamma D_B}{v^2_w} \right)\nonumber\\
k_2 \equiv \frac{v_{w}}{D_L} \left( 1 + \frac{\Gamma D_L}{v^2_w} \right)
\eea
and
\bea
\left(
\ba{c}
a_{11}\\
a_{12}
\ea \;
\right) = \;
\left(
\ba{c}
1\\
\frac{3\Gamma D_B}{v^2_w (D_L/D_B -1 )}
\ea \;
\right) \;\;\;\;, 
\left(
\ba{c}
a_{21}\\
a_{22}
\ea \;
\right) = \;
\left(
\ba{c}
\frac{\Gamma D_L}{v^2_w (D_B/D_L -1 )}\\
1
\ea \;
\right).
\eea
Now, the constants $C_{1,2}$ and $B,L$ can be determined from (\ref{bc1}) and
(\ref{bc2}). In the limit $D_B \ll D_L$ the result for $B$ is
\bea
B = \frac{\Gamma D_L}{v^2_w} n^0.
\label{bden}
\eea
Although the dependence on the wall velocity seems to have a singular limit 
when $v_w \rightarrow 0$ (there is only one power of the wall velocity in 
$n^0$), this is only because we have made the approximation that the
ratio $\frac{\Gamma D_L}{v^2_w} \ll 1$. In the limit $v_w \rightarrow 0$, this 
approximation is obviously not valid and indeed the solution of the diffusion 
equations in this case gives $B=0$, as expected.
  
Finally, in order to compare this result with the experimental one 
$B/s \sim (4$--$6)\times 10^{-11}$, we need to divide by the entropy at the temperature 
of the phase transition, $s =  \frac{2 \pi^2}{45} g_* T^3$, 
where $g_* \sim 110$ counts the degrees of freedom of the relativistic
particles at the electroweak phase transition.
Putting everything together we find generically a effect of the order of
\bea
B/s \sim \frac{\Gamma D_L}{v_w}\; J_{123} \times 10^{-4}.
\eea
If we assume $v(T)/u(T) = v(0)/u(0)$, we can use  
 the experimental bound  $J_{123} < 10^{-7}$ (\ref{eb2}). Considering the 
values
quoted in the literature for the ratio $\Gamma D_L/v_w \sim 10^{-2}$  
(within the SM) \cite{jpt} \cite{vw} \cite{sph}, 
we get a baryon to entropy ratio two orders of magnitude smaller than 
required. 
However, the bounds on $J_{123}$ only hold if the ratio of the scalar vevs 
does not vary with the temperature, which is not necessarily true.
For instance, a variation by a factor of $2$ in the
right direction (i.e. a larger ratio at $T$), increases the result by two
orders of magnitude. This is because the CP asymmetry 
goes like $O(D/S)^6$ which, up to Yukawa couplings, is $\sim O(v(T)/u(T))^6$. 
An enhancement due to this effect has been suggested 
in the context of two-Higgs models in \cite{jpt}.
In order to establish whether this enhancement could take place, a 
detailed study of the scalar potential is required, which is beyond the
scope of this paper.

\section{Conclusions}

We have considered the possibility that 
baryogenesis occurs during the weak phase transition in a
minimal extension of the Standard Model, which contains
extra neutral leptons and conserves total lepton number. 
The leading CP asymmetries come from
the reflection of both neutral and charged leptons
on the bubble wall. Due to the large mean free path of the leptons as compared
to the  typical values of the wall thickness, the calculation is done in the thin wall regime. 
The CP-violating  phases come from two sources. 
For the NHL there are unremovable CP phases due to the non-trivial space 
dependence of the mass matrix inside the bubble wall.
The effect turns out to be tree level and in agreement with naive estimates.
It is only suppressed by the mixing angles.
For the charged leptons there is no tree-level contribution 
and the CP-violating phases appear in the one-loop thermal corrections 
to the lepton propagation in the plasma. The naive estimate gives a 
suppression
in the charged lepton masses and in loop factors ($1/4 \pi$),
besides that in the mixing angles. The result of \cite{fs} suggests that 
the suppression on the charged lepton masses could be absent;
however, we  find agreement with the naive estimate. We
argued that the effect found in \cite{fs} requires a large
flavour dependence of the leading $T$ thermal corrections, which 
is not the case in this type of models.

Using the present constraints on the mixing angles, we
obtain that the leading effect comes from the neutral 
sector and gives
$B/s \leq \frac{\Gamma D_L}{v_w} \;10^{-11}$. 
Assuming SM estimates for the lepton diffusion constant $D_L$,
the wall velocity $v_w$ and the sphaleron rate $\Gamma$, 
we get $B/s \leq 10^{-13}$, which even though the errors
involved are very large, seems too small to account for the 
observed baryon asymmetry. 
However, the constraints on  the mixing angles apply only  if the ratio of the 
scalar vevs at the temperature of the phase transition is the
same as  today, which is not necessarily true.
 It would interesting
to study a realistic scalar potential  to determine whether this
possibility is realized. 

Finally, we want to comment on other scenarios where the baryon asymmetry is 
also generated  
at the electroweak phase transition, through lepton reflection on the bubble 
wall. In ref. 
\cite{cknn} the singlet majoron model was considered. The CP asymmetry 
in that case was also due to the reflection of neutrinos. However, the 
relevant phase space was around the mass of the $\tau$-neutrino 
$\sim O$(10 MeV). Although 
the asymmetry obtained was roughly of the correct order of magnitude, we
think that thermal corrections to the dispersion relation of the 
$\nu_\tau$ in the plasma, which were neglected in \cite{cknn}, should be
taken into account.  In particular, from the calculation of the damping rate of neutrinos
in this model \cite{dampneu}, it is
clear that the typical reflection time of the light 
neutrinos is much larger than the lifetime of 
the quasi-particles in the plasma. In this situation, we expect a considerable
suppression in the CP asymmetry. 
 In refs. \cite{jpt}\cite{cline}, the reflection of 
$\tau$ leptons was considered as the leading contribution to the baryon asymmetry in the two-Higgs model, in the thin wall regime. The  
effects of the damping rate have also been neglected in this case. 
The results for the charged lepton contribution to the asymmetry in the 
present work show that this effect could be important.

\vspace{1cm}

\begin{center}
{\bf Acknowledgements}
\end{center}

We thank G. Anderson,  B. Gavela, M. Joyce, O. P\`ene,
 and D. Tommasini for useful discussions. 
We thank C. Quimbay and S. Vargas-Castrillon for giving us a
program to compute the dispersion relations for charged leptons.
This work was supported in part by CICYT under grant AEN-96/1718
and by DGICYT under grant PB95-1077 (Spain).

\vspace{1cm}

\setcounter{section}{0}
\setcounter{equation}{0}
\renewcommand{\thesection}{Appendix \Alph{section}.}
\renewcommand{\theequation}{A. \arabic{equation}}

\section{}

In the calculation of the CP asymmetry in the neutral sector (section 4)
we have used  the exact propagator in the presence of a wall 
in position space. We give here the expression for $S^{(0)}$
(eq. (\ref{prop})) in the boosted frame, $p_x = p_y =0$: 

\beq
S^{(0)} = S^{(0)}_{left} + S^{(0)}_{across} +  S^{(0)}_{right}
\eeq
with
\bea
S^{(0)}_{left} (z_2,z_1)  \gamma^0
&=& -\Theta(-z_1) \left\{
\Theta(z_2-z_1) \Theta(-z_2) e^{i E (z_2-z_1)} {1+\alpha_z \over 2}
\right. \nonumber \\
&+ & \left. \Theta(z_1-z_2) e^{- i E (z_2-z_1)} {1-\alpha_z \over 2}
+ \Theta(-z_2) e^{-i E (z_2+z_1)} {1-\alpha_z \over 2}
\frac{m \gamma^0}{E+p'} \right\} 
\eea

\bea
S^{(0)}_{across} (z_2,z_1)  \gamma^0
&=& -\Theta(-z_1) \Theta(z_2) e^{-i E z_1} e^{i p' z_2}
\left(1 + \frac {m \gamma^0}{E+p'} \right)  {1+\alpha_z \over 2}
\nonumber \\
& & - \Theta(z_1) \Theta(-z_2) e^{i p' z_1} e^{-i E z_2}
{1-\alpha_z \over 2} \left(1 + \frac {m \gamma^0}{E+p'} \right)
\eea

\bea
S^{(0)}_{right} (z_2,z_1)  \gamma^0
&=& \Theta(z_1) \left\{
-\Theta(z_2-z_1)  e^{i p' (z_2-z_1)}  {1 \over 2}
\left({E \over p'} + \alpha_z + {m \over p'} \gamma^0 \right) 
\right. \nonumber \\                                      
&- & \Theta(z_1-z_2) \Theta(z_2) e^{- i p'(z_2-z_1)} {1 \over 2}
\left({E \over p'} - \alpha_z + {m \over p'} \gamma^0 \right) 
 \nonumber \\                                      
&+& \left. \Theta(z_2) e^{i p'(z_2+z_1)} {1 \over 2}
\left({E \over p'} + \alpha_z + {m \over p'} \gamma^0 \right)                            
\frac{m \gamma^0}{p+p'} \right\}
\eea
and $p'= \sqrt{E^2 - m^2}$.

The propagator in position space for the massless 
left-handed neutrino is
\beq
S^{(0)}_{\nu_L} (z_2,z_1) = \Theta(z_2-z_1) e^{i E (z_2-z_1)}. 
\eeq

\setcounter{equation}{0}
\renewcommand{\theequation}{B. \arabic{equation}}

\section{}

\bea
I_{1a}^i &=& \int_0^\infty dz V(z) e^{i(E+p_i)z}  
=\frac{1}{\delta_H (E+p_i)^2}(e^{i(E+p_i) \delta_H} -1)  \\
I_{2a}^{ij} &=& \int_0^\infty dz_1 \int_0^\infty dz_2 V(z_1) e^{i E z_1} 
e^{i p_i z_2}  \left\{ \Theta(z_1-z_2)e^{ip_j(z_1-z_2)}  
 +\Theta(z_2-z_1)e^{ip_j(z_2-z_1)}-e^{ip_j(z_1+z_2)} \right\}  
\nonumber \\ 
&=& i \frac{2 p_j}{M_i^2-M_j^2} [I_{1a}^i-I_{1a}^j]
\\   
I_{3a}^{ij}&=& \int_0^\infty dz_1 \int_0^\infty dz_2 
\int_{z_2}^\infty dz_3 V(z_1) V(z_2) V(z_3)
e^{iE(z_1-z_2+z_3)} e^{i p_i z_3} \\
& &  
\left\{ (E+p_j) \Theta(z_1-z_2)e^{ip_j(z_1-z_2)}+(E-p_j) 
[\Theta(z_2-z_1)e^{ip_j(z_2-z_1)}-e^{ip_j(z_1+z_2)}] \right\} 
\nonumber \\
I_{5a}^{ijk}&=& \int_0^\infty dz_1 \int_0^\infty dz_2 
\int_{z_2}^\infty dz_3  \int_0^\infty dz_4 \int_{z_4}^\infty dz_5
V(z_1) \ldots V(z_5)
e^{iE(z_1-z_2+z_3-z_4+z_5)} e^{i p_i z_5} \nonumber \\
& &\left\{ (E+p_j) \Theta(z_1-z_2)e^{ip_j(z_1-z_2)}+(E-p_j) 
[\Theta(z_2-z_1)e^{ip_j(z_2-z_1)}-e^{ip_j(z_1+z_2)}] \right\}\nonumber \\
& & 
\left\{ (E+p_k) \Theta(z_3-z_4)e^{ip_k(z_3-z_4)}+(E-p_k) 
[\Theta(z_4-z_3)e^{ip_k(z_4-z_3)}-e^{ip_k(z_3+z_4)}] \right\} 
\eea

\bea
I_{1b}^i &=& I_{1a}^i(-p_i) - I_{1a}^i(p_i)   \\
I_{3b}^{ij} &=& I_{3a}^{ij}(-p_i)-I_{3a}^{ij}(p_i)  \\
%I_{3c}^{ijk} &=& \frac{I_{3b}^{ij}}{2 p_j} - \frac{I_{3b}^{ik}}{2 p_k} \\
I_{5b}^{ijk} &=& I_{5a}^{ijk}(-p_i)-I_{5a}^{ijk}(p_i).
\eea

\end{document}